
\parindent=10pt
\parskip=25pt
\magnification=1200
\baselineskip=13pt

\rightline {KCL-TH-94-4}
\rightline{hep-th/9403185}
\vskip 1in
{\centerline { {\bf{ $W$ Strings and Cohomology in
Parafermionic Theories }}}}
\vskip 1in
{\centerline {\bf{Peter West}}}
\par
{\centerline {Mathematics Department}}
\par
{\centerline {King's College}}
\par
{\centerline {Strand}}
\par
{\centerline {London WC2R 2LS}}
\vskip 1in
{\underbar {Abstract}}
\par
By enforcing locality we relate the cohomology found in parafermionic theories
to that occurring in $W$ strings. This
link provides a systematic method of finding states in the
cohomology of $W_{2,s}$ strings.
\vfil
\eject

\parskip=18pt
The states and scattering amplitudes of the $W_3$ string
contain Ising model primary fields and correlators
respectively [1].  The Ising model that appears in this
way is of a specific type, namely the two scalar
parafermionic representation [2].  However, the
representation of
these parafermions was not the standard one [3,5], but it
was discovered [4] that there exists a field
redefinition which relates the parafermions that occur in
the $W_3$ string to the usual ones.  In the notation of
reference [4] the $W_3$ string is constructed from the
fields $\phi_1, \phi_2$ and $x^\mu$; we will use the
notation of this paper, with the exception that we
replace $\phi_1$ by $\hat
\phi_1$.  Parafermions occur in all critical $W$ strings
; for a $W$ string is based on an algebra whose maximal
spin is $k + 1$, the relevant parafermions are [2,4].
$$
\eqalignno{
\psi_1 &= \exp (i \hat \phi_1 - {\sqrt {k + 2\over
k}} \phi_2)\cr
\psi_{-1} &= -{1\over k}\{i(k + 1) \partial^2
\hat \phi_1 + (k + 1) (\partial \hat \phi_1)^2\cr
&+ i {\sqrt {k (k + 2)}} \partial \hat \phi_1
\partial \phi_2 \} \exp (-i \hat \phi_1 + {\sqrt {k +
2\over k}} \phi_2) &(1)\cr}
$$
The variables $\hat \phi_1$ and $\phi_2$ are linearly
related to the fields $\varphi$ and $\rho$ that occur in
most of the literature on the $W_3$ string by the
relation $\hat \phi_1 = - (k+1) \rho -i {\sqrt {k(k+2)}}
\varphi,\ \phi_2 = -i {\sqrt {k(k+2)}}\rho +
(k+1)\varphi$.  The case
$ k = 2$
corresponds to the $W_3$ string and $W_N$ strings
correspond to taking $N = k + 1$.
\par
The usual parafermions discussed
extensively in the literature [3,5] are
$$\eqalignno{
\psi_1 &= -i \partial \phi_1 \exp \left(-i \phi_1 -
{\sqrt {{k +2\over k}}} \phi_2\right)\cr
\psi_{-1} &= -{1\over k} \left(i(k + 1)
\partial \phi_1 + {\sqrt { k (k +2)}} \partial
\phi_2\right)\cr
&\qquad \exp (i \phi_1 + {\sqrt {{k +2 \over
k}}}\phi_2 ). &(2) \cr}
$$
We will refer to $ \phi_1$ and $\phi_2$ as the
original
variables and $\hat\phi_1$ and $\phi_2$ as $W$ variables.
The relation between the $W$ variables and
the original variables being [4]
$$
e^{i\phi_1} = i \partial \hat \phi_1 e^{-i \hat
\phi_1},\ \  - i \partial \phi_1 e^{-i\phi_1} = e^{i\hat
\phi_1} \eqno (3)
$$
\par
Parafermions are well known to possess three screening
charges [5] which commute with $T(z)$ and $\psi_{\pm 1}
(z)$.
These are given by
$$
Q_i = \oint d z j_i (z) \qquad i = 1,2,3 \eqno (4)
$$
where
$$
\eqalignno{
j_1 &= \exp (i(k + 1) \phi_1 + {\sqrt {k (k
+2) }} \phi_2) \cr
j_2 &= \exp i \phi_1\cr
j_3 &= -i \partial \phi_1 \exp (i \phi_1 - {\sqrt
{k\over k + 2}} \phi_2) &(5)\cr }
$$
In terms of $W_3$ variables two of the screening charges
for $k = 2$ are recognisable as charges used in the
$W_3$ string.  The charge $Q_3$ is the screening charge
$S$
of reference [6] which was used to construct the infinite
number of states that occur in the BRST cohomology of the
$W_3$ string, while $Q_1$ is a nilpotent part of the
BRST charge $Q = Q_0 + Q_1$ of the $W_3$ string.
To make this latter identification one must use the
formula [4]
$$
e^{i n \phi_1} = (-1)^{n+1} e^{-(n+1) \hat \phi_1}
\partial^n e^{i\hat \phi_1} \eqno (6)
$$
The third screening charge $Q_2$ becomes a total
derivative under the change to $W$ variables and so, at
first sight, appears to play no role.
\par
It was also found [2,4] that all the states of the $W_3$
string could also be generated by the action of
parafermions on four basic physical states.  This
construction of the physical states and the above
identification of $Q_1$ leads one to hope that one could
exploit the simplicity of $Q_1$ in terms of original
variables to solve the cohomology of $W$ string theories.
  However, the cohomology of
$Q_1$ is trivial [7].  This is most easily seen by
considering the variables
$$
j_1 (z) = \eta(z), \quad \xi (z) = \exp \{ -i (k +
1) \phi_1 - {\sqrt {k (k+2)}} \phi_2 \}. \eqno
(7)
$$
It is straight forward to verify that the fields $\eta$
and $\xi$ have conformal weights 1 and 0 respectively and
operator product expansion $\eta (z) \xi (w) = {1\over
(z-w)} = -\xi (w) \eta (w)$.  In other words, they have
the same properties as spin 1 and 0 fermionic ghosts.  As
usual, for such a system [8], we define vacua $\mid \pm
>$
to obey $\eta_0 \mid-> = 0$ and $\xi_0 \mid +> = 0$ and
so
$\eta_0 \mid + > = \mid - >$.  Since $Q_1 = \eta_0$,
$Q_1$
vanishing on
any state means that the state possess only the $\mid -
>$
vacuum. However, this vacuum can be written as
 $Q_1$ acting on a $\mid + >$ vacuum.
Introducing the
fields $\eta^\prime = e^{i \phi_1}, \xi^\prime = e^{-i
\phi_1}$ we conclude, using the same argument, that the
cohomology of $Q_2$ is also trivial.
\par
This result is somewhat puzzling when one recalls, that
for the cohomology of the $W_3$ string, the operator $Q_1$
plays an essential role.  The key to understanding this
apparent paradox is that when calculating a cohomology
one must specify not only the BRST charge, but also the
set
of allowable states or operators.  An example of where
this is
important is the Ramond sector of the superstring where
one of the zero modes of the bosonic ghosts is excluded
from the space of allowed states.  In terms of $W$ variables a
minimum requirement is that the physical states be
single-valued.  This means that states
which contain terms with $\phi_1$ and $\phi_2$ without
derivatives and not in the exponential are excluded.
Examining the transformation of equation (3) between the
original variables and the $W$ variables one finds that
although $e^{i\phi_1}$ and $\partial e ^{-i \phi_1}$ lead
to well defined operators, many operators, such as
 $e^{-i m \phi_1}, m = 1,2,3,\dots$, do not.  As such, when
wishing to consider the cohomology of $Q_1$ relevant to
the $W$ string we must restrict ourselves to operators which
are well defined in terms of $W$ variables and so
generated from the operators $e^{i \phi_1}, \partial e^{-i
\phi_1}$ in the sense of including all operators contained
in the repeated operator product expansions of these
operators.  This construction includes
$e^{in\phi_1};
n = 0, 1, 2, \dots,$ but excludes $e^{-i m\phi_1}; m = 0,
1, 2, \dots.$  The charge $Q_2$ is readily found to
commute with $e^{i\phi_1}$ and $\partial e ^{-i \phi}$
and hence all well defined operators.  Thus the criterion
for operators when written in terms of original variables to
be well defined in terms of $W$ variables is that
they commute with $Q_2$.  Since $\phi_2$ is unchanged by
the
transition between the two variables, it does not occur
in the criterion for being well defined.  One can arrive
at this result another way, the operator
$Q_2$ in terms of $W$ variables takes the form $\oint
dz \ \partial (e^{-i \hat \phi_1})$ which is a total
derivative.  Consequently, acting on any state which is
well defined it must vanish.
\par
The above discussion means that when working in terms of
original variables we should consider the ``well
defined'' cohomology of $Q_1$ which are states which are
well defined in the above sense and in the cohomology of
$Q_1$.  Such a state, $\psi$ must satisfy
$$
Q_1 \psi = Q_2 \psi = 0 \eqno (8)
$$
and we regard two states $\psi^\prime$ and $\psi$ as
being equivalent if they differ by $Q_1 \wedge
$ where $\wedge$ is a well defined state, ie $ Q_2 \wedge
=
0$. Since the cohomology of $Q_2$ is trivial this means
two equivalent states should differ by $Q_1 Q_2 \wedge$.
Another way
of stating this result is that the ``well defined''
cohomology of $Q_1$ is given by
$$
{\ker Q_1 \cap \ker Q_2\over Im \ Q_1Q_2} \equiv
H^{W D}_{Q_1} \eqno (9)
$$

We have assumed that $Q_1$ and $Q_2$ have well defined
action on all the states we wish to consider.  This
property is true if we consider Fock spaces based on the
exponentials
$$
\exp (i n \phi_1 + {q\over {\sqrt{k (k +2)}}}
\phi_2) \eqno (10)
$$
for $n,q\ \epsilon \ {\bf Z}$ and we restrict our space
of
states to be of this form.  In terms of the $W$ string
variables
$\rho$ and $\varphi$ this restriction means the states
can be written in terms of the ghosts $d = e^{-i \rho}$
and $e = e^{i \rho}$ and
have $\varphi$ momenta quantized in units of ${i\over
{\sqrt {k (k +2)}}}$.  It will often prove
useful to reparameterize this exponential in terms of
$\ell$ and $m \quad \ell, m\  \epsilon\  {\bf Z}, \ \ell
-
m \ \epsilon \ 2 {\bf Z}$ where $n = {\ell - m\over 2}, q
=
({\ell -m\over 2}) k - m$.  We refer to space of oscillator
states based on such an exponential as $F^{\ell}_m$ and
the exponential itself are $\phi^\ell_m$.
\par
The cohomology of equation (9) is precisely that
discussed in
reference [7] where it was argued that it coincided with
the usual
cohomology studied in parafermionic theories.  The latter
is usually
formulated in terms of the charge $Q_3 =S$ and it is known to
consist of
descendants of the parafermionic primary fields which are
not
themselves parafermionic highest weight states.
It was shown in reference [7]
that for the
cohomology of equation (9) there existed an
isomorphism
between the cohomology of $F^\ell_m$ and
$F^\ell_{m+2nk}$.  As
such we can restrict our attention to $\mid
m\mid\le k$.
\par
The parafermions $\psi_{+1}$ and $\psi_{-1}$ can be
written in the form of a commutator with $Q_2$ and $Q_1$
respectively. Using these results, the operator
product expansion $\psi_1 (z) \psi_p(w) = (z -w)^{2p\over
k} c_{1, p} \psi_{p+1} (w)$, and the
analogous result for $\psi_{-p}$, one finds [15] that
$$
\psi_p\ \propto [Q_2, \ \phi^{-(k+2)}_{-k+2p}], \ \psi_{-
p} \propto[
Q_1, \phi^{-(k+2)}_{k-2p}] \eqno (11)
$$
Evaluating these expressions yields the results
$$
\eqalignno{
\psi_p \ \propto \ &\left (e^{-i \phi_1} {\partial^p\over
p !} e^{i  \phi_1} \right ) e^{-i p \phi_1 - p \phi_2
{\sqrt {k+2\over k}}} \cr
&= {(-1)^{p+1}\over p !} e^{i p \hat \phi_1 - p \phi_2{\sqrt
{k+2\over k}}}\cr
\psi_{-p} &\propto \left(e^{-i\phi^\prime_1} {\partial^p
\over
p !} e^{i \phi^\prime_1} \right) e^{i p \phi_1 + p {\sqrt
{k+2\over k}} \phi_2} & (12)
\cr}$$
The fields $\psi_k$ and $\psi_{-k}$ have conformal weight
zero and for any field $L (z)$ we obtain a new field
$ (\psi_k L) z = \oint d\ w \ {\psi_k (w)\over (w-z)}
L(z)$ and similarly for $\psi_{-k}$. Using
 equation (12) one can
show that the operator product expansion $\psi_k
(z) \ \psi_{-k} (w)$ does not contain any singular terms
as $z \to w$, but does contain the identity.  As such we
may act with $\psi_{-k}$ to invert the effect of $\psi_k$
or visa versa.  This implements the isomorphism $F^\ell_m
\to F^\ell_{m + 2 k}$ discussed above.  In fact, one can
find expressions that write $\psi_p$ and $\psi_{-p}$ as
commutators of $Q_1$ and $Q_2$ respectively.
\par
To explore which states are in $H^{W.D}_{Q_1}$ let us
consider the
field
$$
Y^{k+1}(z)= e^{(-i(k + 2)\phi_1-
\sqrt {k(k +2)}\phi_2(z))}.\eqno (13)
$$
which has a conformal weight $k +1$.  Taking its
commutation
with $Q_2$ we find
$$[Q_2,Y^{k+1}(z)]= {1\over (k + 1)!} e^{-i(k +2)\phi_1-
\sqrt{k(k +2)}\phi_2}\partial^{k+1} e^{i\phi_1}$$
$$=\oint{d\varepsilon\over\varepsilon^{k+2}}
\exp\{-i(k+2)\phi_1(z)-\sqrt{k
(k+2)}\phi_2(z)+i\phi_1(z+\varepsilon )\} \eqno (14)$$
Taking the commutator with $Q_1$, and expanding the
bracket
with the
$\varepsilon$ inside it one finds that
$$
U^{k +1}(z)={1\over (k + 1)!}\sum\limits^{k +1}_{p=0}(-
1)^p({k+1\choose p})^2\biggl[e^{-
i\phi^\prime_1}\partial^p
e^{i\phi_1^\prime}\biggr]\biggl[e^{-
i\phi_1}\partial^{k +1-p}e^{i\phi_1}\biggr] \eqno (15)
$$
where $\phi^\prime_1=(k +1)\phi_1-i\sqrt {k(k+2)}\phi_2$.
In fact $U^{k +1}$ is the spin
$k+1$ primary field that occurs in $\psi_1(z)\psi_{-1}(w)$.
 This follows from the observation that
$[Q_2,Y^{k+1}(z)]=e^{-i\rho} = d$.
\par
Given any state $\psi \in H^{W.D.}_{Q_1}$ then
$U_0^{k +1}\psi =
Q_1Q_2Y_0^{k +1}\psi$ and consequently unless $\psi$,
which can
be chosen to be an eigenvalue of $U_0^{k +1}$, has
eigenvalue
zero it will be in the trivial equivalence class of
$H^{W.D}_{Q_1}$.  If we
restrict our attention to states $\psi$ which are
exponentials $\phi^\ell_m$ of
the form of equation (9) then they are annihilated by
$Q_1$ and
$Q_2$ if $\ell\geq\mid m\mid$.  Their eigenvalue of
$U_0^{k
+1}$ is found from
$$
U^{k +1}_0\phi^\ell_m (z)=\oint\limits_z (w-z)^k
U^{k+1}(w)\phi^\ell_m (z)
$$
$$
\eqalignno{=&\sum\limits_p(-1)^p{k+1\choose p}\oint
{d\tau\over\tau^{p+1}}\oint
{d\varepsilon\over\varepsilon^{k
+2-p}}\oint\limits_z dw (w-z)^{k-n(k +1)+q-
n}\cr&(w+\tau-z)^{n(k +1)-q}(w+\varepsilon-z)^n
\exp\biggl(-i\phi^\prime_1(w)+i\phi_1^\prime(w+\tau)\cr
&-i\phi_1(w)+i\phi_1(w+\varepsilon)+in\phi_1(z)+{q\over
\sqrt{k (k +2)}}\phi_2 (z)\biggr)\cr
&=\lambda\phi^\ell_m (z)& (16) \cr}$$
where
$$
\lambda=\sum\limits^{k +1}_{p=0}(-1)^p{k+1\choose p}
{{\ell +m\over2 }\choose p}{{\ell-m\over 2}\choose k+1-p}
\eqno (17)
$$
The final step is achieved by expanding the brackets
containing
$\tau$ and $\varepsilon$, carrying out the $w$ integral
and observing which terms do not vanish.
\par
Clearly, every term in the above sum vanishes if
$p >{\ell +m\over
2}$ and $(k +1-p)> {\ell -m\over 2}$ which imply
$k
+1 > \ell$.  As such exponentials for which $\ell\le
k$ and $\mid m\mid\le\ell$ have eigenvalue zero and can
be
non-trivial
elements of $H^{W.D.}_{Q_1}$.  We have not carried out a
systematic
search for other solutions, but low level cases suggest
that for
$\mid m\mid\le \ell$ these are the only solutions.  We
now assume that there are no other solutions.
Since $Q_1$ and $Q_2$
commute
with the parafermions, an element of $F^\ell _m$
constructed by the action of the parafermionic modes will
be trivial if $\phi^\ell_m$ is trivial. As such, cohomologies based on all
other exponentials which do not have eigenvalue zero will be trivial.
It may be possible to extend this result to
the full Fock space. The above argument, when taken with the results of
reference [7], allow one to conclude that the cohomology of equation (9)
is the same as the more usual cohomology, which is based on $S$,
studied in parafermionic theories.
\par
The result of the above discussion is that to find non-trivial
elements of $H^{W.D}_{Q_1}$ we must consider Fock
spaces based on
$$
\phi^\ell_m=\exp\biggl({i\over 2}(\ell -m)\phi_1+{\phi_2\over
2\sqrt {k (k+2}}(\ell k - m (k+2))\biggr) \eqno (18)
$$
$\ell =0,1,...,k\ \ ;\ \ m=-\ell ,-\ell +2,....\ell$
These are the
parafermionic primary fields which have conformal
dimensions $\triangle^\ell_m = {\ell (\ell+2)\over
4(k +2)} - {m^2\over 4k}$. It is also useful to consider the
fields
$\hat\phi ^\ell_m = A_{{m-2 \over k}-1} \dots A_{{\ell +
2\over k} -1} A _{{\ell \over k} -1} \phi^\ell_\ell, \ \
m=\ell,\ell +2,...,2k -\ell$  which  have conformal
dimensions $\triangle^{k-\ell}_{k-m} = {\ell (\ell
+2)\over 4(k +2)} - {m^2\over 4k} + {m - \ell
\over 2}$.   In particular, one finds that
$\hat\phi^0_{2p} \propto\psi _p$. In terms of $W$ string variables,
the $\phi^\ell_m$ fields become
$$
\phi^\ell_m = (-1)^{{\ell -m\over 2}} \exp \left(-i
({\ell -m \over 2} +1)\hat \phi_1\right) \partial ^{{\ell
-m\over 2}} \exp (i \hat \phi_1)\quad \exp {\phi_2\over
2{\sqrt{k (k +2)}}}
\left(\ell k - m(k + 2)\right) \eqno (19)
$$
upon use of equation (3).
\par
It was observed in reference [18] that the unitary minimal series for the
$W_n$ algebra contained as its first member a theory with central charge
$c= {2(n-1) \over n+2}$ which coincided with that for the parafermionic theory
with $k=n$ and that the former theory  possesses highest weight
fields, with respect to the $W_n$ algebra, with dimensions equal to the spin
and thermal operators of the
parafermionic theory.
The parafermions are known [23] to generate in their operator
product expansion $\psi_1(z) \psi _{-1}(w)$ a
$W$ algebra which is made up from  a $W_k$ algebra and an infinite number of
higher spin, but null generators. Since the fields $\phi^l_m$  are annihilated
by $A_{{m \over k}+p} ,\ p\ge 1$ and $A_{{-m \over k}+p}^{\dag} ,\ p\ge 1$
they will be annihilated by all the positive modes of the $W$ algebra.
As such the
  $\phi^l_m$ will be highest weight fields with respect to the $W_n$ algebra.
The same applies to
parafermionic descendants which are themselves
parafermionic primary fields.
As such, the states of the parafermionic theory carry a
representation of the $W_k$ algebra. The uniqueness of the $W$ unitary series
then allows one to conclude that the parafermionic theory contains the highest
weight  states and descendants of the first member of the $W_k$ minimal series.
 Checking  the dimensions of the operators that
appear in the two theories for low values of $k$ one indeed finds  that there
is
a one to one match of the  conformal dimensions of the fields of the
two theories. The operator product suggests that all  bilinear parafermionic
excitations $\psi_1\psi_{-1}$ can be written in terms $W$ descendants and
the null nature of the higher spin generators suggests that the
irreducible representation contains only the descendants of the $W_k$ algebra.
Thus, it seems likely that the parafermionic states can be identified with
those
of the first member of the minimal series.

  Since the $W$ generators are found in the operator product of the
parafermions
it follows that they commute with $Q_1$ and $Q_2$. Consequently, we can act
with them on any element of $H^{W.D}_{Q_{1}}$ and create another element.
{}From the arguments given above it is clear that states which can be
written
in terms of the spin $k+1$ $W$ generator are cohomologically trivial and
the case of the parafermions with $k=2$  suggests that the same applies
to all the infinite number of higher spin $W$ generators.
However, this statement does not apply, in general,  to states which can
be written in terms of the generators of the $W_k$ algebra. Thus it
 would seem
likely that the cohomology of equation (9), is none other than the states of
the first member of the minimal series, in agreement with the statement at
the end of the above paragraph.

Let us consider a BRST charge $Q$ of the form
$$
Q = Q_0 + Q_1 \eqno (20)
$$
where
$$
Q_0 = \oint d z c\{ T^{\phi_1, \phi_2} + T^{x} -
{1\over 2} T^{b,c}\} \eqno (21)
$$
For the case of $k=2$ this is the BRST charge of the
$W_3$
string constructed by Thiery Mieg [10] and written in the
above form in reference [4].  In the above system of
fields we have two types of ghost numbers corresponding
to the assignments $(0,-1),\ (0,1)\ (-1,0)$ and $(1,0)$
for the ghosts $b, c, d$ and $e$ respectively.  With this
choice, $Q_1$
and $Q_0$ have ghost numbers $(1,0)$ and $(0,1)$
respectively and consequently, $Q_0^2 = Q^2_1 = \{Q_0,
Q_1\} = 0$.  For $k \ge 3$ the charge of equation (20) is
the BRST charge for the so called $W_{ 2, s}$ strings
discussed in reference [16].  While for $k = 2$ it is the
BRST charge for the $W_3$ string
\par
The charge $Q_0$ can be regarded as that for a critical
string constructed from $\phi_1, \phi_2, x^\mu$ with the
usual ghosts $b,c$.  Although three of the former fields
possess background charges we can, by a linear
transformation, rotate the background charge into one
direction without affecting the cohomology.  The physical
states for such a system were found [12] to be
constructed from D.D.F. like operators in the same way as
for the bosonic string without a background charge.  It
would then seem most likely that one could apply the
arguments of reference [13] to conclude that the
cohomology of $Q$ contained operators of the form
$$
cR(\phi_1, \phi_2, x^\mu) \eqno (22)
$$
where $R$ is a conformal operator of weight one, the
identity operator, 1 corresponding to the $S L
(2, {\bf R})$ vacuum state, and the conjugates of these
states, namely $\partial c c R(\phi_1, \phi_2, x^\mu)$
and
$\partial^2 c \partial c c $.
\par
Given $\chi = \sum \limits_{p^\prime + q^\prime = r}
\chi^{(p^\prime,q^\prime)}$, the equation $Q \chi = 0$
can be separated into a number of equations, one for each ghost number
pair $(p^\prime, q^\prime)$.  For the highest $q$ ghost
number in $\chi,\  Q_0\chi^{(p,q)} = 0$ and the lower ghost
number equations are $Q_1 \ \chi^{(p,q)} + Q_0
\chi^{(p+1, q-1)} = 0$ etc.  This situation is just that
required to apply the theory of the spectral sequence of
a double complex to deduce the cohomology [14].  In this
method one constructs a
series of spaces $E^{(p^\prime,q^\prime)}_r$ and operators $d_r\ : \
E_r^{(p^\prime,q^\prime)}\to E^{p^\prime + r, q^\prime-r+1}_r$
which eventually
terminate in the sense that $d_s$ vanishes or
$E^{(p^\prime, q^\prime)}_{s +1} = E^{(p^\prime,
q^\prime)}_s$.  The
cohomology of $Q$ is then $H^t_Q = \sum\limits_{p^\prime
+q^\prime = t}
E_s^{(p^\prime,q^\prime)}$.  Applying this method to our
case, we have
$E_1 = H_{Q_0}$ and $E_2 = H_{Q_1}( H_{Q_0})$ with $d_1 =
Q_1$.
\par
Given a BRST operator $Q$ and operator $L_n \equiv \{
b_n, Q \}$ which obeys the Virasoro algebra one can
choose the non-trivial cohomology classes to be
constructed from primary fields of weight zero with
respect to $L_n$.  Given a $\psi$ such that $Q\psi = 0$,
then
$L_0 \psi = Q \{ b_0 \psi\}$ and consequently unless $L_0
\psi = 0, \psi$ is in the trivial cohomology class [13].
However, even if $L_n \psi, n\ge 1$ is non-vanishing we
can systematically add terms of the form $Q\wedge$ such
that $L_n \psi = 0 \ \ n\ge 1$.  This is easily achieved
at low levels, indeed at the lowest level one must add
${1\over 2} L_{-1} L_{-1} \psi = Q ({1\over 2} b_{-1} L_1
\psi)$.  The result to all orders follows from writing
$\psi = \psi + (\wp -1) \psi$ where $\wp$ is the physical
state projection operator of reference [17].  The above
result applies to the ordinary bosonic string, the superstring
 and $W$ strings.
\par
Carrying out the first step of the spectral sequence, we
consider a non-trivial cohomology class of $Q_0$ as our
highest $q$ ghost number state.  When $q=1$, this
state is of the form of equation (22), we have $\chi = cR
+ U +$ terms of lower $q$ ghost number, where $R$ and $U$
are conformal fields of weight 1 and 0 respectively.
Applying $Q \chi = 0$ we find $Q_1 c R = -Q_0 U$ which
implies that
$$
Q_1 R = \partial U \eqno (23)
$$
This equation in turn implies that $Q_1 U = 0$.  In
principle, even though $U$  has $q$ ghost number zero
it could contain a $bc$ term.  However, an examination
of this possibility  shows that
such a term must have zero coefficient and we believe this
applies to all terms with more than one $b$.
Since $Q_1U= 0$ the sequence of equations terminates.
\par
Applying a similar analysis to the case where the highest
ghost number in $\chi$ has $q = 1$ ie $\partial c c R$
one finds that $Q_1 R = 0$.  The identity state and its
conjugate are automatically annihilated by $Q_0$ and
$Q_1$ and so are also members of the cohomology of $H_Q$
by themselves.
In terms of the spectral sequence method the above
implies that the operator $d_2$ which maps $\chi^{p,q}$
to $Q_1 \chi^{p+1, q-2}$ vanishes and so the cohomology
of $Q$ is given by $H_Q = H_{Q_1} H_{Q_0}$.
\par
Clearly, we can choose $U$ to be the identity operator;
$\phi^0_0 =1$ in
which case $Q_1 R = 0$.  In this case the physical states
are annihilated by $Q_1$ and $Q_0$ and are a product of a
operator function of $\phi_1$ and $\phi_2$ and one of
$x^\mu$.  States of this type
are of the form
$$
c \phi^\ell_m \vee^x_{1-\triangle^\ell_m}\eqno (24)
$$
where $\vee^x_{1- \triangle^\ell_m}$ is a conformal field
constructed from
$x^\mu$ of weight $1-\triangle^\ell_m$. Clearly, for every
non trivial element of  $H^{W.D}_{Q_1}$ of fixed
conformal weight we can find a
physical state. As discussed above, we will, in general, find
$W_k$ or parafermionic descendants of $\phi^\ell_m$ which are
elements of  $H^{W.D}_{Q_1}$ and so they can be used to
construct physical states. All these
 states have an $x^\mu$ momentum
that is not fixed and are sometimes called
continuous momentum states. For the cases of $k=3,4$ and $5$
we can recover the low level states found in reference
[16] using mathematica.
\par
Other physical states can be found by choosing other
elements $U$ in $H^{W.D}_{Q_1}$ which have conformal
weight $0$.  Examining the weights of $\phi^\ell_m$ and
$\hat \phi^\ell_m$, we find the only possible solutions
are $\phi^k_k$ and $\hat \phi^0_{2k} = \psi_k$. We
also consider $\psi_{-k}$. A
solution to $Q_1 R = L_{-1} U$ is found in each case by
considering $R$ to be proportional to $d_{-1} \phi^k_k$
and $d_{-1} \psi_{\pm k}$.  In the former case, working in terms
of $W$ variables we find that
$$
d_{-1} \phi^k_k = e^{i(k + 1) \hat \phi_1 - {\sqrt
{k\over k+2}} (k+3) \phi_2} \eqno (25)
$$
and
$$\eqalignno{
&Q_1(d_{-1} \phi^k_k) (z) = \oint {d\varepsilon \over
\varepsilon^{k+2}} \oint \limits_z dw (1 +
{\varepsilon\over w-z})^{k+1} (w-z)^{k+1}\cr
&\exp \{-i \hat \phi_1(w) + i \hat \phi_1 (w+\varepsilon)
- i (k+1)\hat \phi_1 (w) + {\sqrt{ k(k+2)}} \phi_2 (w) \cr
&+ i(k+1) \hat \phi_1 (z) - {\sqrt {k\over k+3}} (k+3)
\phi_2 (z) \} \cr
&= - (k+2) \partial \phi^k_k & (26)\cr}
$$
Similarly, one finds that
$$
Q_1 (d_{-1} \psi_k) (z) =
 - {2k + 1 \choose k} \partial \psi_k, \
 Q_1 (d_{-1} \psi_{-k}) (z) =
 - {2k + 1 \choose k+1} \partial \psi_{-k}.
\eqno (27)
$$
In the above cases, we set $R=d_{-1}U$,  whereupon
 equation (23) becomes
$W_{-1}U \propto L_{-1}U$ which we recognise as a highest weight
descendant of the $W_k$ algebra.
\par
The above states are those in the cohomology of $Q$ subject to
the restriction  $\ell \ge |m|$.
As discussed earlier, we may construct, using the
isomorphism between $F^\ell_m$ and $F^\ell_{m+2k}$, an
infinite number of non-trivial elements of the
cohomology of $Q_1$ and so also $Q$.  Although $\psi _{\pm 1}$
commute with $Q_1$, they do not commute with  $Q$ . However,
we can, using equation (27), extend them to the physical states
$\Psi_{k} \equiv c d_{-1} \psi_{k}+ {2k + 1 \choose k} \psi_k$
 and
$\Psi_{-k} \equiv c d_{-1} \psi_{-k}
+ {2k + 1 \choose k+1}  \psi_{-k}$ which do commute with
$Q$. Thus, given any state $\phi$ in the cohomology of $Q$ we can generate
an infinite number of states given $\Psi_{\pm k}^n \phi$.
For the case of $k=2$, evidence based on mathematica calculations,
was given for the above states.
 One can also use the parafermions
$\psi _{\pm 1}$ or the screening charge $S$ in conjunction with an
appropriate picture changing operator $P$ to generate an infinite
number of physical states. In the latter two methods
 one need work with basic states that are only a subset
of the states included in the above restriction.  These latter
constructions are discussed in more detail below.
\par

We have taken care to impose that the physical states when written
in terms of $W$ variables are local.
 However, it could happen that some of these physical states which
are local in terms of $W$ variables are non-local in terms of
original variables. The criterion that a state when written in terms
of $W$ variables is local in terms of
original variables is that it is annihilated by $Q_w = \oint dz
e^{i\hat \phi}= \oint dz e^{(-i(k+1)\rho + \sqrt {k(k+2)}\varphi)}$.
Since the parafermions and the  $\phi^l_m$ are
 are  well defined in terms of $W$ variables, they are annihilated by $Q_w$.
Consequently, all the above states and those found using the isomorphism
are annihilated by $Q_w$.  The conjugates of these states
are constructed using not only the parafermions, but also the picture
changing operator. However, the picture changing operator
is not well defined in terms of original variables and so neither are the
conjugate states.

For the $W_3$ string, $k=2$, it was shown in reference [6] that all the
physical states could be constructed from four basic states by applying
the screening charge $S$ and the picture changing operator $P$ in an
appropriate manner. The
 continuous
momentum operators could be found from the basic operators $a(h,0)$ for
$h= 0, {1\over2}$ and${1\over 16}$ in the notation of that reference
,  while for the discrete operators one used the basic operator $D(0)$ of
reference
[4]. An alternative method [4] of finding these operators was found by
repeatedly applying
the parafermion $\psi_{-1}$ to the  operators $a(0,0)$, $\bar a(0,0)$
and $a({1\over 16},0)$. In addition to these operators, the cohomology also
contains the conjugates of the above operators that can be generated by the
action of screening charges or the parafermion $\psi_1$ and picture changing
operator on the basic states[4]. These states were recovered [19] by an
 alternative, but subsequent method,
which is related to the use of screening charges of reference [6].
A number of arguments [12,20,6,4,19] have been given to
suggest that these are all the physical states in the $W_3$ string.

The  parafermionic primary operators $\phi^l_m$ for $k=2$ are given by
$\phi^0_0,\ \phi^1_{1}$,
$ \phi^1_{-1}, \phi^2_2,\ \phi^2_0$ and $ \phi^2_{-2} $.
These are none other than the operators $a(0,1),\ a({1\over 16},1)$,
$a({1\over 16},2),\ \bar a(0,0),\ \bar a({1\over2},0)$ and $\bar a(0,1)$
 respectively. While taking $U$ to be $\psi_2$ and
$\phi^2_2$ leads to the discrete states  $D(0)$ and $\overline
D (0) = S P D (0)$ respectively.
Thus for the $W_3$ string,
the states of equations (24) to (27), and those generated from them by using
the isomorphism
relating $F_m^l$ to $F_{m+2k}^l $, and their conjugates account for all
states in the cohomology of $Q$.
\par
One can repeat the above pattern of the $W_3$ string for the $W_{2,k+1}$
strings. Rather than use the isomorphism generated by $\Psi _{\pm k}$ we can
use the method of the screening charges. One finds that the
action of $S ^n \phi ^\ell _m $ is well defined if
$n{(\ell-n+1)k \over 2(k+2)} -{nm \over 2} \in {\bf Z}$.
It is straight forward to find the solutions to this equation. For example,
one finds that if  $k=3$ then $S ^{5q} \phi ^\ell _m $ and
$S ^{5q +\ell + 1 } \phi ^\ell _m $, $q\in {\bf Z}$ are the allowed states;
while if $k=4$ then $S ^{3q} \phi ^\ell _m $ for $q \in {\bf Z}$  and
$S ^{3q +\ell + 1 } \phi ^\ell _m $ where $q \in {\bf Z}$ if $\ell$ is even and
$q \in {\bf 2Z}$ if $\ell$ is odd. In this  construction one must insert
appropriate picture changing operators to obtain a non zero result.
One can also use the action of the parafermions $\psi _{\pm 1}$ to create new
states.
Although conceptually  simpler, it is more difficult to find explicit
expressions for the states using this method. Unlike the $W_3$ string, we do
not
expect to get all states in this construction unless, as discussed, above,
we must take account the $W_k$ or parafermionic descendants. One could also
generate states using the screening charge $Q_w$ and the picture changing
operator which was used for the case of $k=2$ in reference [16].
\par
It should be possible to extend
the techniques used in this paper to find not only all the physical
states in the
$W_{2,s}$ strings,
but also those for the $W$ strings. This could be achieved by exploiting
the knowledge of  $H^{W.D}_{Q_1} $ to solve the full cohomology of $Q$.
This is the strategy advocated in reference [24] for the case of $W_4$ strings.
\par
Under the change of variables $\phi_1 \to
\phi^\prime_1, \  \phi_2 \to \phi^\prime _2 = -i {\sqrt
{k(k+2)}} \phi_1 - (k+1) \phi_2$ one finds that the energy
momentum tensor maintains its form and so $Q_0 \to
Q_0^\prime, \ \psi_1 \to k\psi^\prime_{-1}, \ \psi_{-1}
\to {1\over k} \psi_1$ and $Q_1 \leftrightarrow Q_2$.
Since the cohomology relevant to the $W$ string is given
by equation (9) this change of variables is a symmetry of
the physical states.  It would interesting to see if this
extends to the scattering amplitudes and what is the
significance of this symmetry.
\par
In recent series of papers it has been shown that the ordinary bosonic string
can be embedded in the $N=1,2$ superstring [21], and the $W_3$ string [22].
However, in order to achieve these results, in particular the matching of the
cohomologies, non-local field redefinitions similar to the one of equation (3)
are required. As such, this paper should shed some light on the relations
between these various theories.

\par
{\underbar {Acknowledgement}}
\par
We wish to thank Mike Freeman, Paul Howe and Steven Hwang
for helpful discussions.

\vfil\eject
{\bf{References}}
\parskip=5pt
\item {[1]}For a review see P. West, ``A review of W
Strings'', preprint G\"oteborg - ITP-93-40, Salamfest to
be published by W.S.P.
\item {[2]}M. Freeman and P. West, Phys Lett
{\underbar{B314}} (1993) 320.
\item {[3]} A. Zamolodchikov and V. Fateev, Sov Phys JETP
62 (1985) 215, JETP 63 (1986) 913.\hfil\break
A. Gersasimov, A. Marshakov and A. Morozov, Nucl Phys
{\underbar{B328}} (1989) 664.\hfil\break
K. Ito and Y. Kazama, Mod Phys Lett A5 (1990) 215.
\item {[4]}M. Freeman and P. West ``Parafermions, $W$
Strings and their BRST charges'' KCL-TH-93-14, hep-th
/9312010; Phys Lett B to be published.
\item {[5]} J. Distler and Z. Qiu, Nucl Phys B 336 (1990)
533. \hfil\break
D. Nemeschansky, Nucl Phys B363 (1989) 665.
\item {[6]}M. Freeman and P. West, Int J Mod Phys A8
(1993) 4261.
\item {[7]} T. Jayaraman, K. S. Narain and M. H. Sarmadi,
Nucl Phys {\underbar {B343}} (1990) 418.
\item {[8]} K. Kato and K. Ogawa, Nucl Phys {\underbar
{B212}}, 443 (1983).
\item {[9]} D. Gepner and Z Qiu, Nucl Phys {\underbar
{B285}} (1987) 423. \hfil\break
J. Distter and Z. Qiu, Nucl Phys {\underbar {B336}}
(1990) 533.
\item {[10]} J. Thiery-Mieg, Phys Lett {\underbar{B197}}
(1987) 368.
\item {[11]} H. Lu, C. N. Pope, S. Schrans and X. J.
Wang, Nucl Phys {\underbar {B408}} (1993) 3.
\item {[12]} P. West Int J Mod Phys {\underbar {A8}}
(1993) 2875.
\item {[13]} M. Freeman and D. Olive, Phys Lett
{\underbar {B175}} (1986) 151.
\item {[14]} For a review see; P. Bouwknegt, J. McCarthy,
and K. Pilch, Comm Maths Phys 131 (1990) 125, Progress of
Theoretical Physics Supplement {\underbar {102}} (1990)
67.\hfil\break
R. Bott and L. Tu, Differential Forms in Algebraic
Topology, Springer-Verlag.
\item {[15]} P. Griffin and O. Hernandez ``Structure of
Irreducible SU(2) Parafermion Modules Derived via the
Feigen-Fuchs Construction'' preprint Fermilab - Pub-
90/117-T.
\item {[16]} H. Lu, C. N. Pope and X. J. wang, ``Higher-
spin strings and W minimal models'' hep-th/9308114.
 H.Lu, C. N. Pope, X.J. Wang and S. C. Zhao, "A note on $W_{2,s}$ strings",
preprint hep-th/9402133

\item {[17]} R. Brower and C. Thorn Nucl Phys {\underbar
{B31}} (1971) 163.
\item {[18]} V.A. Fateev and S. L. Lykyanov, Int. J. Mod. Phys. {\underbar
{A3}}
(1998) 507.
\item {[19]} H.Lu, C. N. Pope "The Complete Cohomology the $W_3$ String",
preprint hep-th /9309041
\item {[20]}  M. Freeman and P. West Phys. Lett. {\underbar { B299}} (1993) 30
\item{[21]} N. Berkovits and C. Vafa, ``On the uniqueness of string
theory'',  hep-th/9310129;\hfil\break
 J. M. Figueroa-O'Farrill, ``On the universal string theory'',
preprint, hep-th/9310200.\hfil \break
N. Ohta and J. L. Petersen, "$N=1$ from $N=2$ Superstrings" preprint
hep-th/9312187
\item {[22]} N. Berkovits, M. Freeman and P. West " A $W$-String Realization
of the Bosonic String", KCL-TH-93-15, Phys. Lett. B to appear.
\item {[23]} I. Bakas and E. Kiritsis, Int. J. Mod. Phys. {\underbar {A7}}
(1992) 339;\hfil \break F. Yu and Y-S. Wu, Phys. Rev. let. 68 (1992) 2996.
\item {[24]} E. Bershoeff, H.J. Boonstra, S. Panda and M. de Roo,
" A BRST Analysis of W- symmetries", preprint UG-4/93; \hfil\break
H.J. Boonstra, "BRST Cohomology of the critical $W_4$ string", hep-th/9401164.

\end